%% file: main.tex
\documentclass{article}
\usepackage{iclr2027_conference,times}

\usepackage{graphicx}
\usepackage{booktabs}
\usepackage{multirow}
\usepackage{amsmath}
\usepackage{amssymb}
\usepackage[fleqn,tbtags]{mathtools}
\usepackage{enumitem}
\usepackage{hyphenat}
\usepackage{balance}
\usepackage{microtype}
\usepackage[detect-weight]{siunitx}
\usepackage[table]{xcolor}
\usepackage{algorithm}
\usepackage{algorithmic}
\usepackage{verbatim}
\usepackage{listings}
\usepackage{wrapfig}
\usepackage{capt-of}
\usepackage{placeins}
\usepackage{hyperref}
\usepackage{url}
\usepackage[capitalise]{cleveref}

\graphicspath{{./images/}}

\definecolor{graycolor}{gray}{.9}

\title{Neural Explicit Shape Representation}

\author{Jun-Hao Wang \quad Changan Lu \quad Baoquan Chen \quad
Peng-Shuai Wang\thanks{Corresponding author.} \\
Peking University}

\hypersetup{
  pdftitle={Neural Explicit Shape Representation},
  pdfauthor={Jun-Hao Wang, Changan Lu, Baoquan Chen, Peng-Shuai Wang}
}

\iclrfinalcopy

\begin{document}

\maketitle
\fancyhead{}
\renewcommand{\headrulewidth}{0pt}

\begin{abstract}
\looseness=-1
Existing neural 3D representations separately address continuous surface modeling, efficient image synthesis, or object-specific compression. They do not jointly provide the explicit surface queries required by ray tracing and a structured output space that can be predicted across objects. We introduce NeuOctree, a neural explicit representation that directly predicts whether and where a ray intersects a surface, together with its normal and albedo. NeuOctree stores shared features at the corners of sparse octree voxels and interpolates them into a continuous local field. A lightweight decoder with recurrent intersection-guided sampling refines each query near the predicted surface, enabling direct surface rendering without mesh extraction or texture baking and accommodating open, non-watertight geometry. The same octree structure supports both compact per-object optimization and a shared octree CNN that constructs renderable assets from colored point clouds without per-object fitting. Across seven meshes and four storage budgets, NeuOctree achieves the lowest positional error in 24 of 28 comparisons while retaining competitive normal accuracy, and substantially improves joint geometry--appearance reconstruction over N-BVH and VQAD. The feedforward model constructs a representation in 0.05 seconds and transfers from Objaverse to ShapeNet without retraining. We further demonstrate ray-traced rendering and image-conditioned generation, showing that NeuOctree provides a common output space for neural inference and direct surface rendering. Code is available at \url{https://github.com/octree-nn/neural-octree}.
\end{abstract}

\begin{figure}[H]
  \centering
  \includegraphics[width=\linewidth]{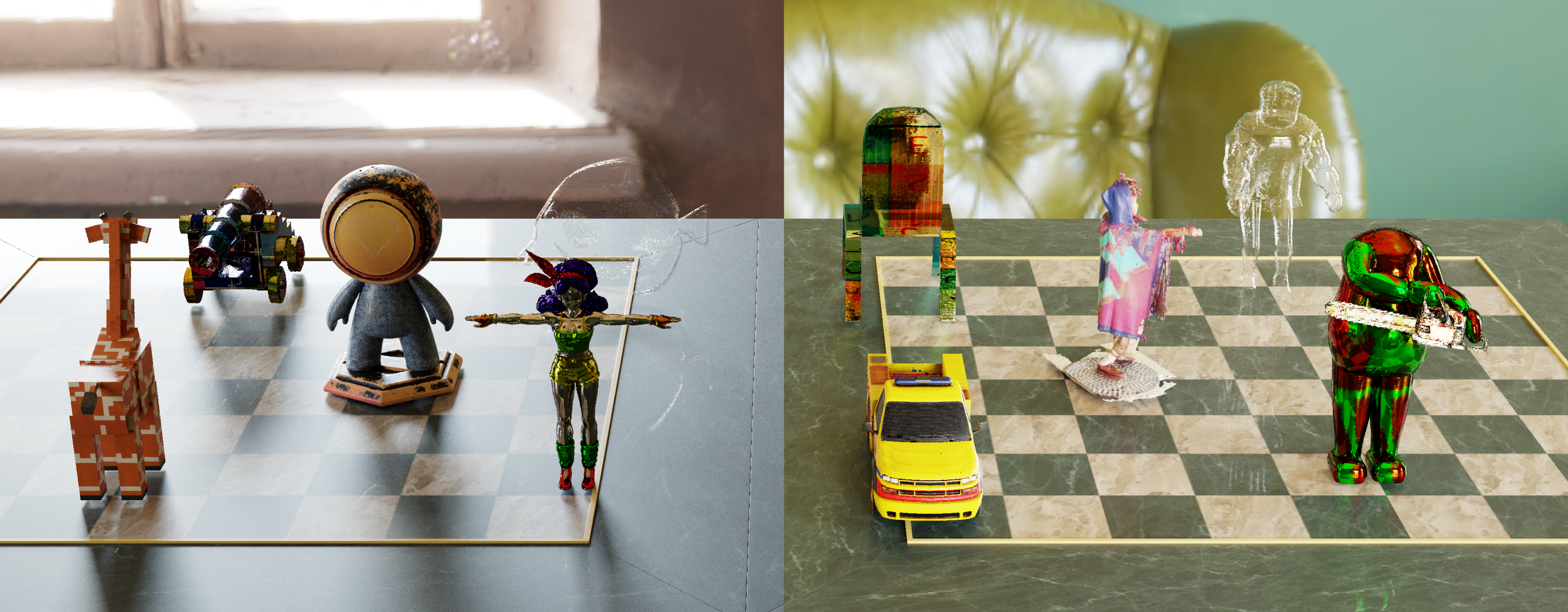}
  \caption{We propose NeuOctree, a novel neural explicit shape representation that directly predicts ray-surface intersections, enabling accurate and efficient ray-traced rendering.}
  \label{fig:teaser}
\end{figure}

\input{src/introduction}

\input{src/relatedwork}
\input{src/method}
\input{src/results}

\FloatBarrier
\input{src/conclusion}

\subsection*{AI use statement}
In this work, we used a generative AI coding assistant to assist with limited code debugging, migrate the existing manuscript into the ICLR 2027 \LaTeX{} template, proofread the language, and check compilation and formatting. It was not used to produce the research ideas, experimental results, figures, or scientific claims in this paper. The authors reviewed the revised source and take responsibility for the final content of this work.

\bibliography{src/ref/reference}
\bibliographystyle{iclr2027_conference}

\clearpage
\appendix
\input{src/appendix}

\end{document}

%% file: src/introduction.tex
\section{Introduction} \label{sec:intro}

Learning compact 3D assets requires a representation that connects neural prediction with geometric computation. A ray-tracing renderer needs to determine where a ray intersects an object and recover the normal and material attributes at that point. A network trained across objects benefits from a structured output space in which to predict geometry and appearance. Bringing these requirements together would allow learned assets to pass directly from a neural network to a ray-tracing renderer.

Existing representations address different parts of this problem. Signed distance fields \citep{Park2019,Chabra2020} and occupancy fields \citep{Mescheder2019} encode surfaces as level sets of learned scalar functions. NeRF \citep{Mildenhall2020} jointly models density and radiance, with hash encodings \citep{Muller2022} and sparse voxel structures \citep{yu2021plenoctrees,Fridovich2022,Liu2020} reducing the cost of field evaluation and rendering. SDF-based volume rendering \citep{Wang2021e} improves surface reconstruction, while Gaussian splatting and sparse voxel rasterization \citep{Huang2024,Sun2025} enable efficient image synthesis. These advances illustrate the value of compact, spatially organized representations. However, density or occupancy values do not directly specify the intersection and material attributes required by a surface ray tracer: these quantities must be recovered through additional surface localization or geometry extraction.
\looseness=-1

Neural compression methods bring learned representations closer to conventional surface rendering. Neural Geometry Fields \citep{Sivaram2024} encode detailed geometry through displacements of coarse mesh patches, while N-BVH \citep{Weier2024n} directly predicts ray intersections and appearance within a bounding volume hierarchy. Both construct their representations through object-specific optimization. Building on the direct ray-query interface, we seek a representation whose geometry and appearance can also be predicted by a shared network across objects.

We introduce \textbf{NeuOctree}, a neural explicit shape representation that couples direct ray--surface intersection prediction with an octree-based sparse voxel structure. The octree organizes local geometry, skips empty space during traversal, and provides a structured domain for 3D neural networks. Adjacent voxels share corner features, defining a continuous feature field through trilinear interpolation. A lightweight, shared MLP decodes the ray and interpolated features into a hit probability, intersection position, surface normal, and albedo. These outputs supply the surface queries needed for ray tracing without mesh extraction or texture baking. Because the representation does not require an inside--outside labeling, it also accommodates open, non-watertight surfaces.

Accurate intersection prediction requires querying the right part of the feature field. Features sampled only at a voxel's boundary may be far from the surface, and rays observing the same surface point can enter the voxel at different locations. We therefore introduce a recurrent sampling strategy that uses predicted intersections to guide subsequent feature queries. This refinement brings queries closer to the surface and encourages rays from different views to access features near the same location. The ablation shows improved intersection and normal accuracy over endpoint-only and uniform sampling.
\looseness=-1

The same structure enables us to amortize representation construction across objects. We build a feedforward 3D U-Net on octree-based CNNs \citep{Wang2017} to predict voxel-corner features from colored point clouds. Averaging predictions at shared vertices produces a consistent feature field, which is decoded by the shared intersection MLP. Training the feature extractor and decoder jointly with rendering losses allows a single checkpoint to construct geometry and appearance for new inputs without per-object optimization. NeuOctree thus supports \textit{both per-object compression and feedforward construction} within the same representation and renderer.

We evaluate per-object compression on seven meshes from Thingi10K and the Stanford 3D Scanning Repository at four storage budgets. NeuOctree achieves the lowest positional error in 24 of 28 model--budget comparisons and competitive normal accuracy compared with geometry-only methods, and significantly overperforms N-BVH and VQAD which jointly represent geometry and appearance. On the separate feedforward task, a shared checkpoint constructs a representation in 0.05 seconds with lower intersection error, compared with approximately 3.5 minutes for N-BVH fitting under the reported settings. The same checkpoint, trained on Objaverse, transfers to ShapeNet without retraining. We further demonstrate direct ray-traced rendering with reshading, reflection, refraction, and fisheye projection, and show image-conditioned generation of NeuOctree assets as a downstream application.
Our contributions are:
\begin{itemize}[leftmargin=*,itemsep=2pt]
    \item[-] \textbf{A structured neural surface representation} that jointly encodes geometry and appearance in shared voxel-corner features and supports direct ray-traced rendering.
    \item[-] \textbf{Recurrent intersection refinement} that adaptively queries the continuous feature field to improve the accuracy of predicted surface intersections and normals.
    \item[-] \textbf{Feedforward representation construction} that predicts renderable NeuOctree assets without per-object optimization, with demonstrated transfer across datasets and native 3D generation as a downstream application.
\end{itemize}

%% file: src/relatedwork.tex
\section{Related Work} \label{sec:related}

\begin{figure*}[t!]
	\centering
	\includegraphics[width=0.95\textwidth]{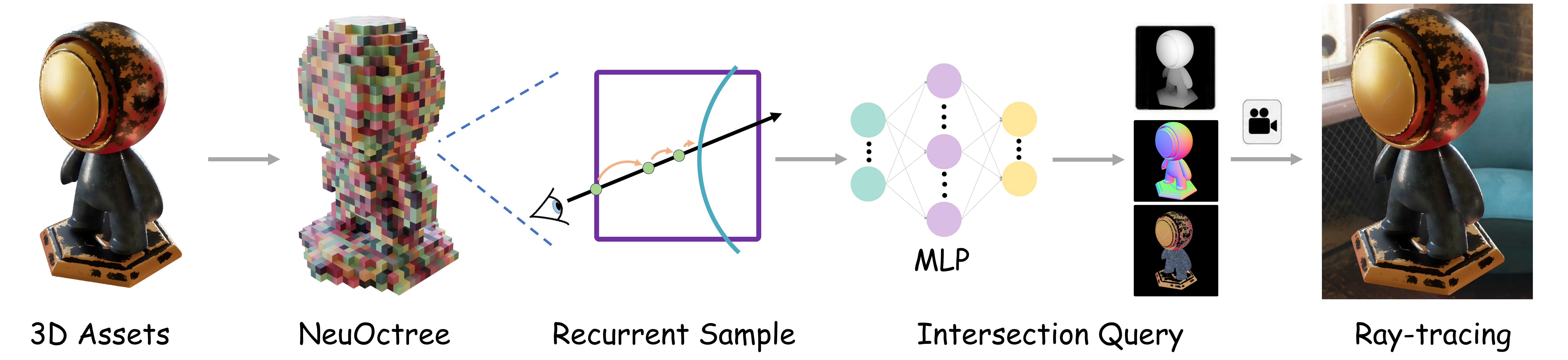} %
	\caption{Overview of the NeuOctree representation and rendering process. The structured representation stores local features in an octree and predicts ray--surface intersections with a shared MLP, enabling direct ray-traced rendering.}
	\label{fig:overview}
\end{figure*}

\paragraph{Neural Surface and Radiance Representations.}
Neural implicit fields provide a compact and differentiable alternative to explicit geometry. Early methods represented shapes with coordinate-based multilayer perceptrons, using signed distance functions (SDFs) \citep{Park2019,Chabra2020} or occupancy fields \citep{Mescheder2019}. Subsequent approaches introduced octrees and adaptive spatial partitions \citep{Takikawa2021,Martel2021}, while neural radiance fields jointly modeled density and appearance through volume rendering \citep{Mildenhall2020}. Octree and hash-grid variants \citep{yu2021plenoctrees,Fridovich2022,Liu2020,Muller2022}, SDF--density hybrids \citep{Wang2021e,Li2023b}, and explicit rasterization-based representations \citep{Huang2024,Sun2025} have substantially improved memory use or image-rendering speed.

These methods nevertheless expose a mismatch with surface ray tracing. A density or occupancy value does not directly specify a ray--surface intersection, normal, or material; obtaining these quantities typically requires repeated field evaluations, surface localization, or mesh extraction. The resulting geometric ambiguity can compromise fine detail, and level-set formulations often rely on an inside--outside convention that is awkward for open or non-watertight surfaces. NeuOctree instead makes the ray--surface query the representation's primary interface: a shared decoder predicts the hit, position, normal, and albedo from a ray and a continuous octree feature field. This design preserves the compactness of neural representations while enabling direct ray-traced rendering without an intermediate mesh or texture-baking stage.

\paragraph{Neural Mesh Compression.}
Traditional mesh compression combines connectivity coding, coordinate quantization, and simplification to reduce storage while preserving surface fidelity \citep{Deering1995,Chow1997,Gutierrez2005,Touma1998,Heckbert1999}. Learning-based methods replace hand-designed operators with graph networks or lightweight MLPs that upsample a coarse mesh and predict vertex displacements \citep{Liu2020c,Chen2023b,Sivaram2024,Pentapati2025}. Although effective for individual assets, these approaches are generally optimized per mesh, focus primarily on geometry, and require a fixed coarse topology; appearance is commonly handled by a separate texture-baking step. N-BVH \citep{Weier2024n} is closer to our setting because it predicts ray intersections and appearance within a hierarchical structure, but it also relies on object-specific optimization.

NeuOctree addresses these limitations with a sparse octree that supplies a global spatial prior and shared voxel-corner features that provide local context. Geometry and appearance are decoded through the same ray-query interface, so the representation can be optimized as a compact per-object asset or predicted by a shared network across objects. The distinction is therefore not only a different compression structure, but also a shift from per-mesh fitting to a renderable representation that supports feedforward construction.

\paragraph{Structured 3D Learning and Feedforward Reconstruction.}
Point-, voxel-, and sparse-voxel networks have become standard tools for 3D perception and generation \citep{Qi2017a,Qi2017,Li2018,Maturana2015,Wu2015,Wang2017,Graham2018,Choy2019,Wang2023,Wu2024}. Their structured spatial domains support scalable feature extraction, but their outputs are typically occupancies, semantic features, or task-specific predictions rather than a representation that can be queried directly by a surface renderer. NeuOctree uses the same structured prior for a different purpose: an octree U-Net predicts features at voxel corners from colored point clouds, and averaging shared vertices yields a consistent continuous field for the ray-intersection decoder. This establishes a direct bridge between feedforward 3D learning and surface rendering: one shared checkpoint constructs geometry and appearance without per-object optimization, while the resulting representation remains suitable for ray tracing and can serve as a native output space for 3D generation.

%% file: src/method.tex
\section{Method} \label{sec:method}

\paragraph{Overview.}
NeuOctree is built around a direct surface query. An asset is represented by an
occupied sparse octree and a continuous feature field defined on its voxel
corners. Given a ray segment inside an occupied voxel, a shared decoder predicts
whether the ray hits the surface and, if so, the intersection position, normal,
and albedo. The same decoder is used for every voxel and every ray; only the
octree topology and its feature values vary across assets. This separation lets
us use the representation in two regimes: a feature table can be optimized for
one asset, or an octree network can predict the feature field in a single
feedforward pass. In both cases the output is directly consumable by a surface
ray tracer and does not require mesh extraction or texture baking.

\subsection{Structured Neural Surface Representation}

\paragraph{Sparse octree and continuous feature field.}
We construct an octree from the input point cloud and retain its occupied leaf
voxels. The octree provides a structured support for spatial queries and lets
the renderer skip empty space. For an occupied voxel $v$, let
$\{\mathbf{f}_{v,ijk}\}_{i,j,k\in\{0,1\}}$ denote its eight corner features.
At a local coordinate $\mathbf{q}\in[0,1]^3$, the decoder reads the continuous
feature field by trilinear interpolation,
\[
    \boldsymbol{\phi}_v(\mathbf{q})
    = \sum_{i,j,k\in\{0,1\}}
      w_{ijk}(\mathbf{q})\,\mathbf{f}_{v,ijk},
\]
where $w_{ijk}$ are the standard trilinear weights. Corner features shared by
adjacent voxels are identified as one vertex feature, so interpolation remains
continuous across voxel boundaries. In per-object optimization, the unique
vertex features are learned directly. In the feedforward model, the network
first predicts eight corner slots for each occupied voxel; slots that refer to
the same geometric vertex are averaged before rendering. This deterministic
aggregation converts voxel outputs into the same shared feature field used by
the optimized representation.

For the colored feedforward setting, the interpolated field can be augmented
with a voxel-local residual feature. We optionally use a pooled input-color
prior in the material branch. These terms provide local appearance context
without changing the surface-query interface or shared vertex field.

\paragraph{Ray-query decoder.}
For a ray traversing voxel $v$, we denote the local entry and exit points by
$\mathbf{p}_{\mathrm{in}}$ and $\mathbf{p}_{\mathrm{out}}$, and use
$\mathbf{d}=\operatorname{norm}(\mathbf{p}_{\mathrm{out}}-\mathbf{p}_{\mathrm{in}})$
as the local ray direction. At a query position $\mathbf{q}$, a geometry branch
receives the interpolated feature, the local position, the ray direction, and
the current refinement state. A separate material branch uses the same spatial
feature field and ray information together with the current position and color
state. Both branches are lightweight residual MLPs whose weights are shared
across all voxels and rays. Their outputs are
\[
    (\mathbf{a},\mathbf{z},s_h)=G_\theta(\boldsymbol{\phi}_v(\mathbf{q}),\mathbf{q},\mathbf{d},\text{state}),
    \qquad
    \mathbf{c}=\sigma\!\left(M_\theta(\boldsymbol{\phi}_v(\mathbf{q}),\mathbf{q},\mathbf{d},\text{state})\right),
\]
where $\mathbf{a}\in\mathbb{R}^3$ parameterizes the position, $\mathbf{z}\in\mathbb{R}^3$
is normalized to produce the surface normal, $s_h$ is the hit logit, and
$\sigma$ is the sigmoid function. The predicted hit probability is
$h=\sigma(s_h)$ and the predicted albedo is $\mathbf{c}\in[0,1]^3$.

The position parameterization is ray-local. In the segment-bounded variant
used by the feedforward rendering configuration, the first component of
$\sigma(\mathbf{a})$ gives $\alpha\in[0,1]$ and
\[
    \widehat{\mathbf{p}}
    = \mathbf{p}_{\mathrm{in}}+
      \alpha(\mathbf{p}_{\mathrm{out}}-\mathbf{p}_{\mathrm{in}}).
\]
The per-object geometry compression protocol uses the equivalent ray-projection
parameterization, which projects the predicted local point onto the current ray
line. When the segment-bounded option is enabled, this projection is restricted
to the current segment. Both forms preserve the key property that the decoder predicts a
surface intersection rather than a density level set. Fourier position features
are not used in the reported configurations.

\subsection{Recurrent Intersection Refinement}

Samples taken only at a voxel boundary can be far from the surface, and the
boundary point depends on viewing direction even when rays observe the same
surface point. A single boundary query therefore makes the decoder learn an
unnecessarily view-dependent mapping. We use the predicted intersection to
place later queries in the continuous feature field.

Let $\mathbf{q}_k$ be the $k$-th adaptive query. The decoder produces an
intersection estimate $\widehat{\mathbf{p}}_k$ and the next query is obtained by
projecting this estimate back onto the current ray parameterization,
\[
    \mathbf{q}_{k+1}=\Pi_{\mathrm{ray}}
    (\widehat{\mathbf{p}}_k),
\]
where $\Pi_{\mathrm{ray}}$ denotes ray-line projection or its segment-bounded
variant, depending on the chosen position parameterization.
The recurrent decoder can maintain a chained hidden state across iterations.
Our pooled variant, used for the reported compression and sampling experiments,
encodes the entry and exit points as ordinary members of the feature set,
initializes the first interior query at the segment midpoint, and then adds
further adaptively placed queries from the preceding readout. The per-query
hidden increments are averaged before a shared output head predicts the final
attributes. With three adaptive iterations this gives five feature reads per
voxel segment (entry, exit, and three adaptive queries). The fixed-sampling
ablations use the same decoder modules and replace only the adaptive positions
with predetermined positions.

This construction couples all queries to the same local feature field while
moving later queries toward the surface. It therefore improves localization
and normal prediction without introducing a large per-voxel network or requiring
volume samples along the entire ray. When the chained recurrent variant is
used, intermediate predictions are also supervised; at inference we use the
final stage.

\subsection{Octree Traversal and Surface Rendering}

Given a camera ray, we enumerate its intersections with occupied leaf voxels in
front-to-back order. Each candidate is represented by its local entry and exit
points and the index of its voxel feature block. The decoder is evaluated on
these candidates in chunks. A candidate is accepted when its hit probability
exceeds a threshold; the first accepted candidate for each ray is selected.
Its local position is converted back to octree coordinates by adding the voxel
origin, while the predicted normal and albedo are retained as surface
attributes. Rays for which no candidate is accepted return the background
value. Thus octree traversal replaces the broad volume sampling used by neural
radiance fields, and the selected attributes can be passed directly to a
standard ray-tracing integrator for reshading, reflection, refraction, or other
camera effects.

\subsection{Feedforward Representation Construction}

The structured representation makes feature construction amortizable across
objects. We build an input octree from a colored point cloud and attach point
color, local point offset, and normalized global coordinates to occupied cells.
An octree-based U-Net \citep{Wang2017} processes this sparse hierarchy with octree convolutions,
downsampling and upsampling stages, residual blocks, and skip connections. At
the query depth, it predicts a tensor with eight feature slots per occupied
voxel. Reshaping these slots and averaging all slots that share a geometric
vertex yields the continuous field used by the decoder.

The feature extractor and the ray-query decoder are trained jointly with the
same rendering losses used for per-object optimization. Consequently, a
single checkpoint maps a new colored point cloud to a renderable NeuOctree
without optimizing the object-specific feature table. The resulting field is
also a native output space for the image-conditioned generation pipeline in
Section~\ref{sec:result_applications}.

\subsection{Training Objective}

Training supervision is obtained by unprojecting RGB--D images into camera
rays. For every occupied ray--voxel segment before the ground-truth surface,
we store a hit label, the local intersection position, the oriented normal, and
the albedo. Let $m$ denote the hit label and let $\mathcal{P}$ be the positive
surface pairs. The pointwise objective is
\[
    \mathcal{L}_{\mathrm{PNCH}}
    = \mathcal{L}_{p}(\mathcal{P})
    + \mathcal{L}_{n}(\mathcal{P})
    + \lambda_h\mathcal{L}_{\mathrm{hit}}(h,m)
    + \lambda_c\mathcal{L}_{\mathrm{color}}(\mathbf{c},\mathbf{c}^{*};\mathcal{P}),
\]
where $\mathcal{L}_{p}$ and $\mathcal{L}_{n}$ are position and normal reconstruction
losses, $\mathcal{L}_{\mathrm{hit}}$ is BCE (or its class-balanced variant), and
$\mathcal{L}_{\mathrm{color}}$ is a relative color reconstruction loss. The
geometry terms are masked by the ground-truth hit labels; in the feedforward
configuration their normalization is restricted to positive surface pairs so
that empty ray--voxel segments do not dilute the surface signal.

For colored reconstruction, we supplement the pointwise terms with image-level
normal and albedo SSIM losses, together with optional smoothness, color
constancy, detail-gradient, and hit-topology regularizers. In compact form,
\[
    \mathcal{L}
    = \sum_{k} w_k\left(
        \mathcal{L}_{\mathrm{PNCH}}^{(k)}
        + \lambda_{\mathrm{img}}\mathcal{L}_{\mathrm{img}}^{(k)}
        + \lambda_{\mathrm{topo}}\mathcal{L}_{\mathrm{topo}}^{(k)}
      \right),
    \qquad \sum_k w_k=1,
\]
where $k$ indexes recurrent stages. Image terms are masked by valid surface
pixels when configured, while hit-topology terms penalize holes, floaters, and
boundary inconsistencies. Geometry-only experiments use the corresponding
subset to supervise the renderer.

%% file: src/results.tex
\section{Results} \label{sec:result}
\begin{table}[H]
\centering
\caption{Geometric compression on seven meshes across four payloads.}
\label{tab:mesh_compression}
\scriptsize
\setlength{\tabcolsep}{1.10pt}
\renewcommand{\arraystretch}{1.05}
\begin{tabular}{@{}lccccccccccccccc@{}}
\toprule
\multirow{2}{*}{\textbf{Method}} & \multirow{2}{*}{\textbf{Model Size}} & \multicolumn{2}{c}{Armadillo} & \multicolumn{2}{c}{Buddha} & \multicolumn{2}{c}{Dragon} & \multicolumn{2}{c}{Einstein} & \multicolumn{2}{c}{Lucy} & \multicolumn{2}{c}{Metatron} & \multicolumn{2}{c}{XYZ} \\
\cmidrule(lr){3-4} \cmidrule(lr){5-6} \cmidrule(lr){7-8} \cmidrule(lr){9-10} \cmidrule(lr){11-12} \cmidrule(lr){13-14} \cmidrule(lr){15-16}
& & $d_{pm}$ & $d_{norm}$ & $d_{pm}$ & $d_{norm}$ & $d_{pm}$ & $d_{norm}$ & $d_{pm}$ & $d_{norm}$ & $d_{pm}$ & $d_{norm}$ & $d_{pm}$ & $d_{norm}$ & $d_{pm}$ & $d_{norm}$ \\
\midrule
\multirow{4}{*}{NGF}
& 81~KiB  & 28.161 & 13.01 & 7.697 & \textbf{9.05} & 21.184 & \textbf{17.39} & 9.561 & \textbf{12.65} & 17.375 & \textbf{12.75} & 22.877 & 7.25 & 16.458 & \textbf{14.60} \\
& 122~KiB & 13.477 & \textbf{9.56} & 6.392 & \textbf{7.62} & 14.547 & \textbf{15.07} & 6.215 & \textbf{10.53} & 8.238 & \textbf{9.50} & 17.709 & 5.69 & 13.363 & \textbf{12.65} \\
& 244~KiB & 11.471 & 7.84 & 3.917 & \textbf{5.66} & 12.688 & \textbf{12.75} & 7.410 & 9.33 & 7.931 & \textbf{7.87} & 13.754 & 4.00 & 23.215 & 13.87 \\
& 823~KiB & 6.859 & 5.80 & 3.350 & 4.52 & 9.348 & 9.84 & 4.324 & 7.01 & 7.672 & 6.94 & 11.544 & 3.11 & 9.779 & 9.27 \\
\midrule
\multirow{4}{*}{QNDF}
& 81~KiB  & 33.591 & 20.54 & 90.764 & 38.03 & 89.450 & 35.54 & 27.726 & 25.08 & 28.939 & 21.97 & 68.320 & 23.92 & 40.575 & 27.70 \\
& 122~KiB & 14.007 & 14.15 & 28.003 & 21.31 & 38.895 & 24.80 & 11.027 & 16.95 & 13.762 & 14.50 & 15.071 & 6.37 & 18.855 & 18.23 \\
& 244~KiB & 7.474 & 9.95 & 5.451 & 9.94 & 17.228 & 17.38 & 5.682 & 12.11 & 5.601 & \textbf{9.39} & 4.975 & 2.96 & 8.169 & 12.07 \\
& 823~KiB & 2.969 & 5.97 & 1.819 & 5.87 & 6.405 & 11.11 & \textbf{1.759} & 7.86 & \textbf{2.276} & 6.34 & \textbf{1.773} & \textbf{1.65} & \textbf{2.791} & \textbf{7.48} \\
\midrule
\multirow{4}{*}{VQAD}
& 81~KiB  & 111.632 & 23.20 & 83.563 & 22.19 & 126.049 & 29.36 & 85.346 & 22.64 & 112.090 & 28.68 & 173.517 & 26.23 & 147.619 & 29.93 \\
& 122~KiB & 98.542 & 21.45 & 53.994 & 18.68 & 96.217 & 27.41 & 48.289 & 18.69 & 74.050 & 24.03 & 139.993 & 20.47 & 108.956 & 26.79 \\
& 244~KiB & 70.747 & 18.09 & 36.205 & 14.14 & 66.713 & 22.37 & 32.017 & 14.27 & 50.375 & 19.01 & 73.962 & 15.43 & 70.348 & 21.84 \\
& 823~KiB & 44.176 & 13.64 & 34.747 & 12.75 & 46.077 & 17.95 & 34.300 & 13.66 & 48.232 & 17.62 & 59.661 & 12.69 & 61.810 & 20.09 \\
\midrule
\multirow{4}{*}{N-BVH}
& 81~KiB & 80.580 & 18.31 & 56.645 & 19.77 & 88.037 & 27.72 & 66.767 & 23.75 & 60.925 & 24.31 & 156.881 & 27.14 & 87.812 & 24.67 \\
& 122~KiB & 79.107 & 16.30 & 52.843 & 18.13 & 77.018 & 25.29 & 67.596 & 23.83 & 56.094 & 23.03 & 138.813 & 24.69 & 81.085 & 22.51 \\
& 244~KiB & 73.780 & 14.04 & 46.715 & 16.31 & 72.329 & 23.27 & 60.489 & 21.38 & 52.143 & 20.52 & 112.875 & 19.41 & 74.981 & 20.30 \\
& 823~KiB & 77.783 & 11.84 & 43.673 & 13.08 & 66.946 & 20.92 & 60.068 & 19.38 & 50.551 & 19.89 & 88.852 & 16.04 & 71.500 & 18.46 \\
\midrule
\multirow{4}{*}{Ours}
& 81~KiB & \textbf{9.616} & \textbf{12.01} & \textbf{6.583} & 11.97 & \textbf{12.542} & 20.18 & \textbf{7.050} & 15.48 & \textbf{9.319} & 14.81 & \textbf{13.867} & \textbf{5.92} & \textbf{12.289} & 16.90 \\
& 122~KiB & \textbf{6.963} & 9.81 & \textbf{5.279} & 9.10 & \textbf{9.521} & 16.99 & \textbf{5.175} & 12.65 & \textbf{7.650} & 12.09 & \textbf{9.845} & \textbf{4.25} & \textbf{9.390} & 14.60 \\
& 244~KiB & \textbf{4.363} & \textbf{7.40} & \textbf{2.719} & 6.73 & \textbf{6.591} & 12.83 & \textbf{3.680} & \textbf{9.24} & \textbf{3.446} & 9.44 & \textbf{3.750} & \textbf{2.68} & \textbf{4.660} & \textbf{11.28} \\
& 823~KiB & \textbf{2.316} & \textbf{4.87} & \textbf{1.685} & \textbf{4.27} & \textbf{3.035} & \textbf{8.26} & 1.821 & \textbf{6.59} & 2.435 & \textbf{5.92} & 2.676 & 2.46 & 3.409 & 8.82 \\
\bottomrule
\end{tabular}%
\end{table}

\begin{table}[H]
\centering
\begin{minipage}[t]{0.48\linewidth}
\centering
\captionof{table}{Joint geometry and appearance on 30 Objaverse and 30 ShapeNet assets.}
\label{tab:feedforward_nbvh}
\scriptsize
\setlength{\tabcolsep}{1.0pt}
\renewcommand{\arraystretch}{1.05}
\begin{tabular}{@{}lrrrrrr@{}}
\toprule
\multirow{2}{*}{\textbf{Method}} & \multicolumn{3}{c}{\textbf{Objaverse}} & \multicolumn{3}{c}{\textbf{ShapeNet}} \\
\cmidrule(lr){2-4} \cmidrule(lr){5-7}
& $d_{pm}$ & $d_{norm}$ & PSNR & $d_{pm}$ & $d_{norm}$ & PSNR \\
\midrule
N-BVH (823~KiB) & 56.235 & 11.842 & 23.641 & 54.284 & 12.567 & 17.375 \\
VQAD ($\leq$823~KiB) & 15.777 & 4.989 & 32.568 & 39.126 & 11.969 & 25.548 \\
Ours ($\leq$823~KiB) & \textbf{2.292} & \textbf{3.104} & \textbf{33.682} & \textbf{1.628} & 8.064 & \textbf{34.877} \\
Ours-F & 4.550 & 6.706 & 28.539 & 5.969 & \textbf{5.229} & 27.834 \\
\bottomrule
\end{tabular}
\end{minipage}\hfill
\begin{minipage}[t]{0.48\linewidth}
\centering
\captionof{table}{Mean fitting time (minutes) over seven meshes at each payload.}
\label{tab:optimization_time}
\scriptsize
\setlength{\tabcolsep}{2.0pt}
\renewcommand{\arraystretch}{1.05}
\begin{tabular}{@{}lrrrr@{}}
\toprule
\textbf{Method} & \textbf{81 KiB} & \textbf{122 KiB} & \textbf{244 KiB} & \textbf{823 KiB} \\
\midrule
NGF   & 3.9  & 4.1  & 6.3  & 19.1 \\
QNDF  & 14.5 & 41.0 & 42.4 & 53.2 \\
VQAD  & 16.0 & 35.2 & 36.7 & 51.8 \\
N-BVH & 3.5  & 3.7  & 3.6  & 3.1 \\
Ours  & 28.2 & 30.2 & 31.1 & 37.7 \\
\bottomrule
\end{tabular}
\end{minipage}
\end{table}

\subsection{Results of Per-Mesh Compression} \label{sec:result_per_mesh}
We compare NeuOctree with the per-object optimization methods NGF
\citep{Sivaram2024}, QNDF \citep{Pentapati2025}, VQAD
\citep{Takikawa2022VQAD}, and N-BVH \citep{Weier2024n} over a broad range of
storage budgets. VQAD is adapted to predict surface normals and intersection
positions under the same geometry supervision.

\paragraph{Dataset and Metrics.} We evaluate the seven meshes in
\cref{tab:mesh_compression} from the Stanford 3D Scanning Repository
\citep{Stanford3DScanningRepository} and Thingi10K \citep{Thingi10K}, rendering
16 fixed $512\times512$ views. We report mean ray--surface position error
$d_{pm}$ in normalized object coordinates, scaled by $10^3$, and oriented normal
error $d_{norm}$ in degrees.

\begin{figure}[H]
    \centering
    \includegraphics[width=\linewidth]{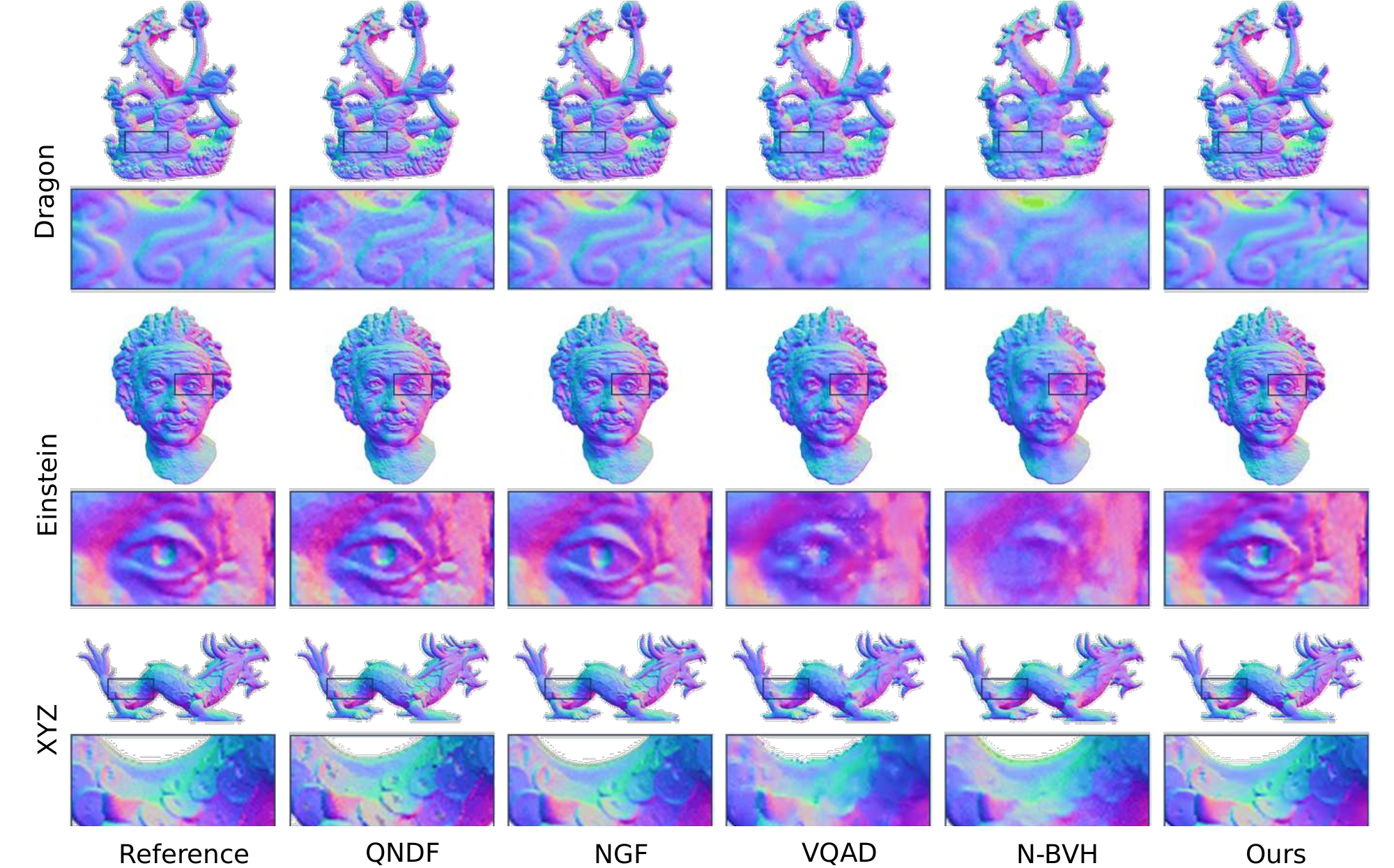}
    \caption{Qualitative compression at 823~KiB on Dragon, Einstein, and XYZ-Dragon.}
    \label{fig:results}
\end{figure}

\begin{figure}[H]
    \centering
    \includegraphics[width=\linewidth]{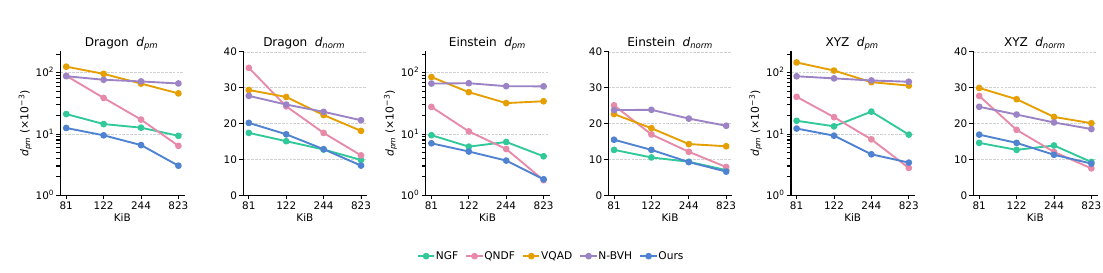}
    \caption{Compression-rate comparison across the three representative meshes in \cref{fig:results}.}
    \label{fig:compression_rate}
\end{figure}

\paragraph{Geometry Compression.} As shown in
\cref{tab:mesh_compression}, NeuOctree obtains the lowest $d_{pm}$ in 24 of 28
model--budget comparisons. It wins all 21 at 81--244~KiB and three of seven at
823~KiB, where QNDF takes the other four. NeuOctree also wins 12 normal-error
comparisons; NGF leads mainly at the smallest payload, while QNDF is competitive
only at the largest.
Thus, the strongest baseline changes with payload, whereas NeuOctree maintains
the most consistent positional accuracy. The curves in
\cref{fig:compression_rate} and qualitative results in \cref{fig:results}
corroborate this trend, showing cleaner silhouettes and better preservation of
fine structures.

\begin{figure}[H]
    \centering
    \includegraphics[width=\linewidth]{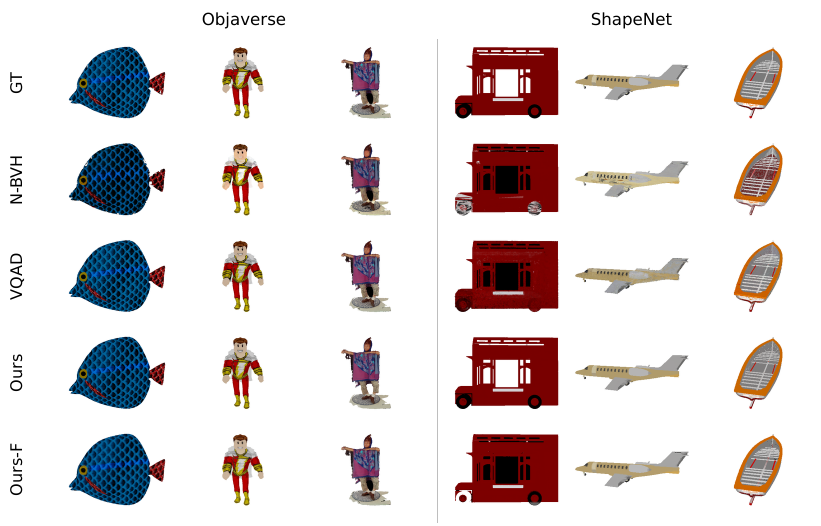}
    \caption{Albedo reconstruction on representative Objaverse and ShapeNet assets.}
    \label{fig:with_color}
\end{figure}

\paragraph{Joint Geometry and Appearance Compression.} On Objaverse,
per-object NeuOctree achieves the best normal accuracy and PSNR; on ShapeNet,
it is best on all three metrics in \cref{tab:feedforward_nbvh}. Relative to
VQAD, it reduces $d_{pm}$ and $d_{norm}$ by 85.5\% and 37.8\% on Objaverse
while adding 1.11~dB PSNR. The ShapeNet improvements are 95.8\%, 32.6\%, and
9.33~dB, respectively. Representative albedo reconstructions are shown in
\cref{fig:with_color}.

\begin{wraptable}[9]{r}{0.48\linewidth}
\vspace{-10pt}
\centering
\caption{Recurrent ray-sampling ablation.}
\label{tab:sampling_ablation}
\small
\setlength{\tabcolsep}{4.5pt}
\renewcommand{\arraystretch}{1.05}
\begin{tabular}{@{}lcrr@{}}
\toprule
\textbf{Sampling strategy} & \textbf{Points} & $d_{pm}$ & $d_{norm}$ \\
\midrule
Endpoints only  & 2 & 3.178 & 7.798 \\
Uniform         & 5 & 3.036 & 7.102 \\
Recurrent       & 5 & \textbf{2.797} & \textbf{6.475} \\
\bottomrule
\end{tabular}
\vspace{-8pt}
\end{wraptable}

\paragraph{Sampling Strategy Ablation.} We compare recurrent sampling with endpoint-only
and four-sample uniform variants. Recurrent sampling reduces $d_{pm}$ and
$d_{norm}$ by 7.7\% and 9.4\% relative to uniform sampling, and by 12.0\% and
17.0\% relative to endpoint-only sampling. This supports querying the feature
field near the predicted surface rather than at fixed locations.

\begin{wrapfigure}[19]{R}{0.46\linewidth}
    \vspace{-8pt}
    \centering
    \includegraphics[width=\linewidth]{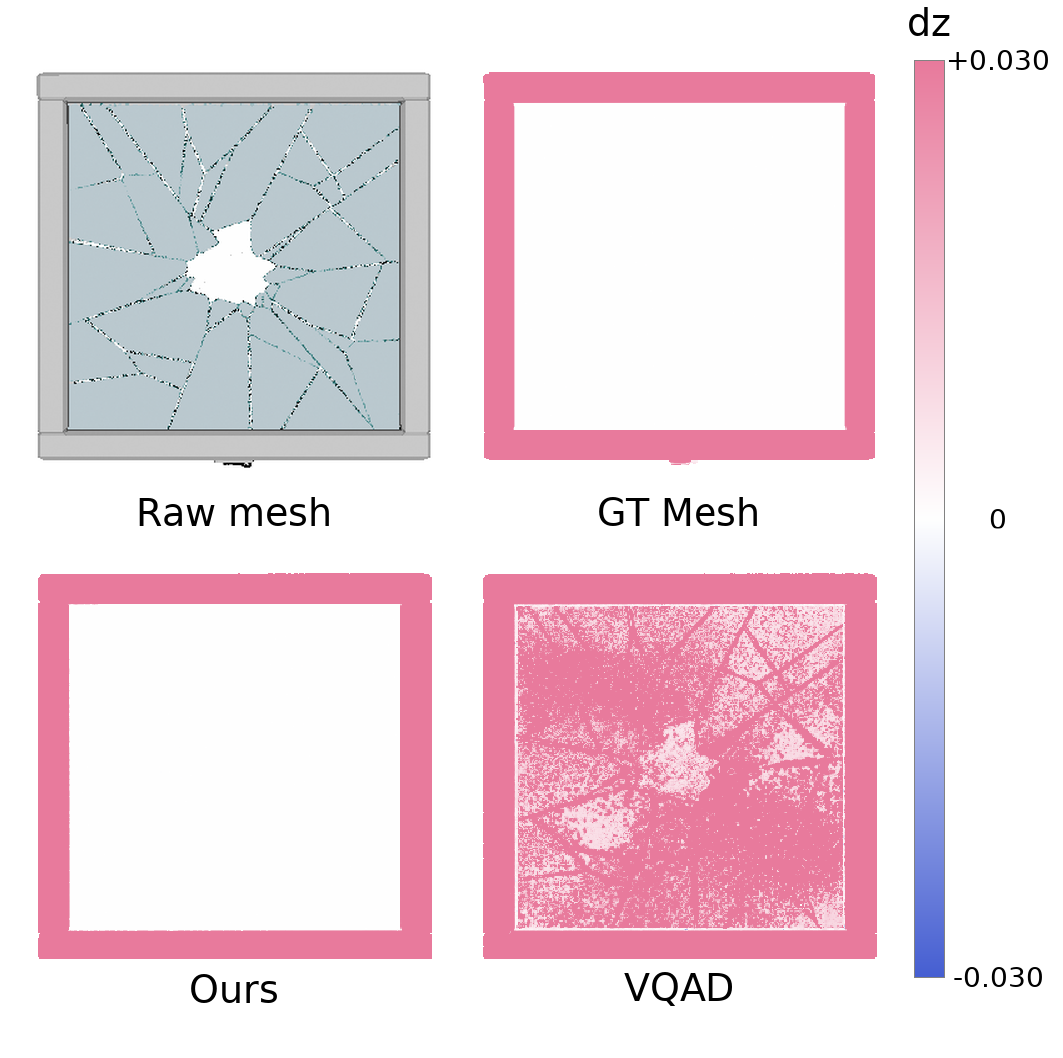}
    \caption{Open pane: two-sided depth consistency.}
    \label{fig:open_surface}
    \vspace{-8pt}
\end{wrapfigure}

\paragraph{Open-Surface Reconstruction.} We evaluate NeuOctree and VQAD on an
open glass pane using an 823~KiB budget. For matched
mirrored face-on cameras, we back-project both depth maps and measure their
separation $dz$ along the pane normal at pixels visible from both sides
(\cref{fig:open_surface}). NeuOctree matches the GT median $dz=0$, whereas
VQAD gives $dz=0.059$ world units, revealing a systematic two-sided offset.
These results support NeuOctree's ability to represent open surfaces without an inside--outside labeling like VQAD.

\subsection{Results of Feedforward Methods} \label{sec:result_feedforward}

\paragraph{Dataset and Metrics.} Our feedforward model is trained on a randomly sampled subset of Objaverse with 32k assets. We evaluate on 30 unseen Objaverse assets and 30 ShapeNet assets to compare with per-object methods. The same metrics $d_{pm}$, $d_{norm}$ as in \cref{sec:result_per_mesh} are used, along with PSNR for appearance reconstruction.

\paragraph{Evaluation.} We evaluate a shared feedforward checkpoint under the
same rendering and scoring protocol used for the per-object methods. Ours-F
retains the direct ray-query interface without per-object fitting. On Objaverse,
its reconstruction is comparable to VQAD: it has substantially lower positional
error, while VQAD remains stronger on normal accuracy and PSNR. On ShapeNet,
Ours-F improves all three reported metrics over VQAD, showing that the same
structured output space remains effective after the dataset shift. 

\paragraph{Efficiency.} \Cref{tab:optimization_time} reports practical fitting
times of each method. N-BVH requires about 3 minutes and VQAD requires
tens of minutes of optimization for each object. In contrast, our feedforward
model constructs a renderable representation from a colored point cloud in
50~ms with 19.5 hours for training paid once across all objects. As shown in \cref{tab:feedforward_nbvh}, this
removes per-object optimization while retaining competitive reconstruction
quality.

\paragraph{Generalization.} The shared model is trained on Objaverse and
evaluated on ShapeNet without retraining. Despite this distribution shift,
our feedforward model outperforms both VQAD and N-BVH on ShapeNet, although both baselines
optimize each object separately. This result shows that the octree feature
extractor and intersection decoder transfer across datasets while preserving the
directly renderable output space.

\subsection{Applications} \label{sec:result_applications}
\begin{figure}[H]
    \centering
    \includegraphics[width=\linewidth]{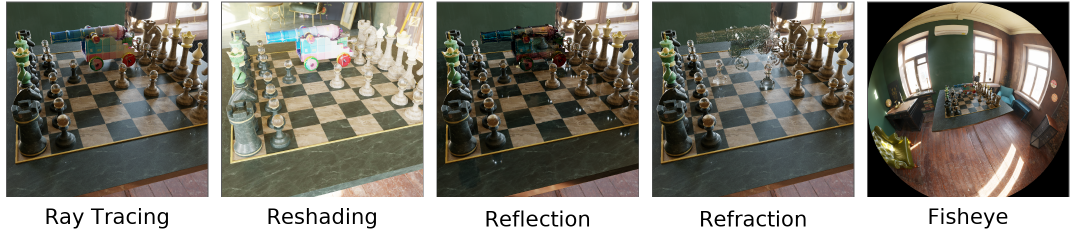}
    \caption{Ray-traced rendering and alternative effects from NeuOctree.}
    \label{fig:ray_tracing}
\end{figure}

\paragraph{Photorealistic Rendering with Ray Tracing.} Since the decoder
returns a hit, position, normal, and albedo, NeuOctree can be queried directly
by a surface ray tracer. It supports reshading, reflection, refraction, and a
fisheye projection without mesh extraction or texture baking
(\cref{fig:ray_tracing}).

\begin{figure}[H]
    \centering
    \includegraphics[width=\linewidth]{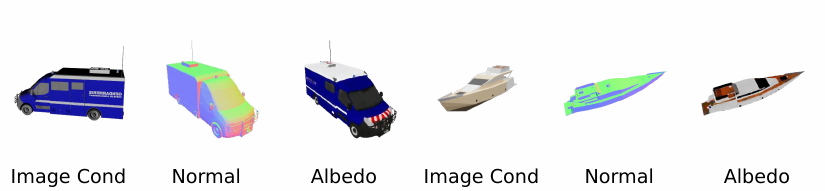}
    \caption{Image-conditioned generation; each group shows the image condition, normal, and albedo.}
    \label{fig:generation_examples}
\end{figure}

\paragraph{Native 3D Generation.} NeuOctree can also serve as the native output
of an image-conditioned generator. A sparse-octree VAE compresses node
features, and a flow-matching DiT predicts occupancy and latent features that
decode directly into renderable geometry and appearance. Each example in
\cref{fig:generation_examples} shows the conditioning image, surface normal,
and albedo.

%% file: src/conclusion.tex
\section{Conclusion} \label{sec:conclusion}
We presented NeuOctree, a neural explicit surface representation for ray tracing.
Shared corner features in a sparse octree and a recurrent decoder jointly predict
ray--surface hits, positions, normals, and albedo, supporting compact
geometry--appearance modeling, direct rendering, and open surfaces. Experiments
show accurate reconstruction across meshes and budgets; the shared feedforward
model constructs a renderable representation in about 50~ms, rather than minutes
of per-object optimization, and transfers from Objaverse to ShapeNet without
retraining. Ray-traced rendering and image-conditioned generation further
demonstrate the versatility of this structured output space.

NeuOctree still requires costly high-fidelity per-object fitting and models albedo
rather than view-dependent materials; generation remains a proof of concept.
Future work will accelerate octree construction, extend surface queries to richer
materials, lighting, and dynamics, and develop NeuOctree as a native output space
for 3D generative models.

%% file: src/appendix.tex
\section{Experimental Details and Additional Results}
\label{sec:appendix}

This appendix records the configurations that determine the entries in the
main paper, with particular emphasis on the capacity-matched comparison in
\cref{tab:mesh_compression}. We separate the per-object optimization setting
from the shared feedforward model because their stored state and training
protocols are fundamentally different.

\subsection{Common Evaluation Protocol}
\label{app:evaluation_protocol}

\begin{wrapfigure}[16]{r}{0.42\linewidth}
\vspace{-8pt}
\centering
\includegraphics[width=0.96\linewidth]{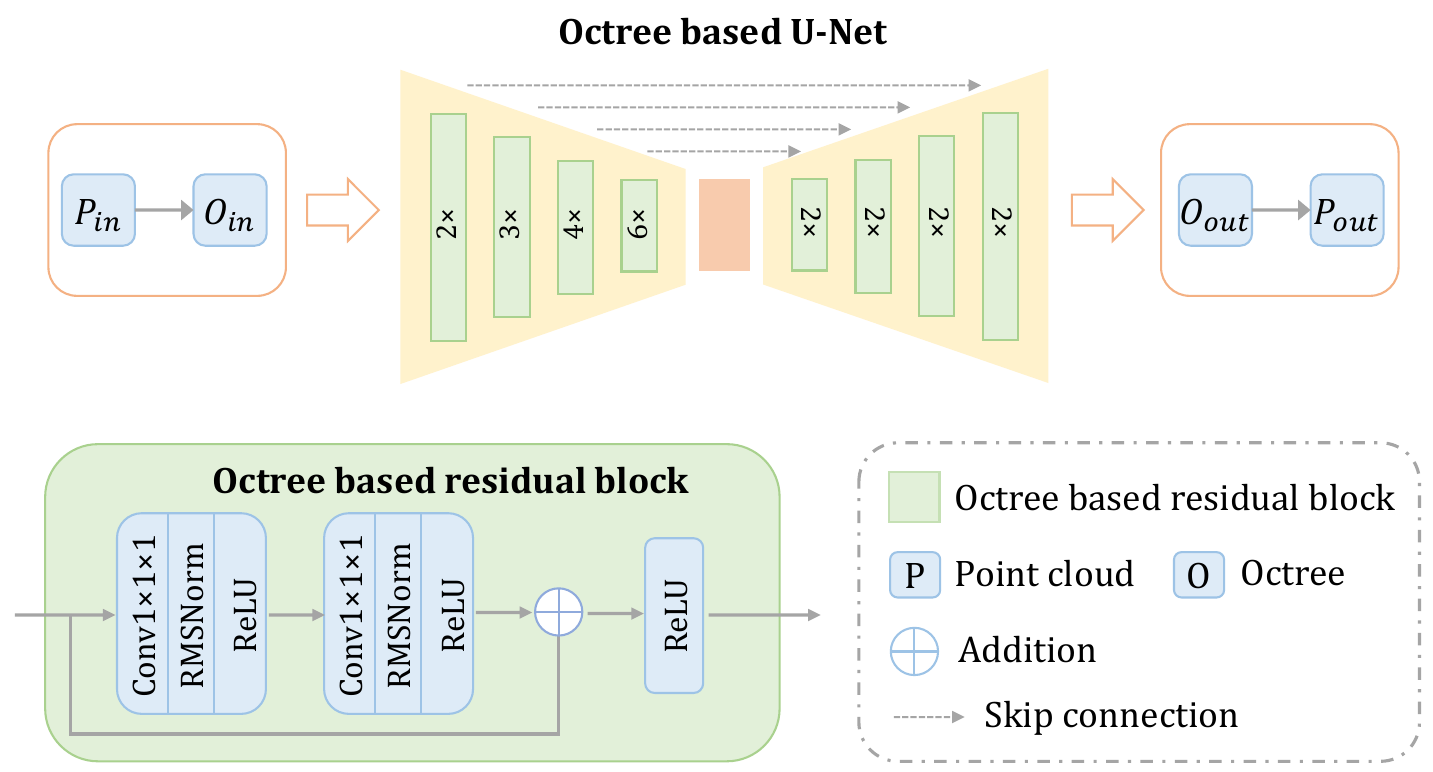}
\caption{3D U-Net used by the feedforward encoder. Multiscale octree
features are decoded through skip connections and octree residual blocks.}
\label{fig:app_ff_unet}
\vspace{-8pt}
\end{wrapfigure}

\paragraph{Objects and cameras.}
The geometry-compression benchmark contains Armadillo, Buddha, Dragon,
Einstein, Lucy, Metatron, and XYZ-Dragon. Every method is scored from the same
16 fixed cameras at $512\times512$ resolution, with a $60^\circ$ field of view
and a camera distance of 2.5 after normalizing the longest bounding-box edge to
2. We use the same per-object coordinate convention and camera conversion for
all methods.

\paragraph{Metrics.}
The evaluation mask is the union of the ground-truth and predicted foreground
masks. The position metric $d_{pm}$ is the mean Euclidean distance between
corresponding ray intersections in normalized object coordinates and is
reported after multiplication by $10^3$. The normal metric $d_{norm}$ is the
mean oriented angular error in degrees. Appearance experiments additionally
report RGB PSNR. Some evaluation implementations store $d_{pm}$ after multiplication
by $10^4$; those values are divided by ten before insertion into the paper.

\paragraph{Payload accounting.}
The four nominal targets are 82,944, 124,928, 249,856, and 842,752 bytes,
reported as 81, 122, 244, and 823~KiB. A payload includes all object-specific
state required by a method at inference time. Shared weights of the
feedforward model are excluded from its per-object representation size. The
tables below record the actual configuration and measured payload used for
each of the 28 cells of every method. We use the abbreviations $D$ (octree
depth), $F$ (feature dimension or requested/coarse faces, as specified by the
caption), $H$ (decoder width), $L$ (octree level), $B$ (codebook bit width),
$Q$ (coarse quads), $N$ (BVH nodes), and $P$ (measured payload in bytes).

\subsection{Feedforward NeuOctree}
\label{app:feedforward_details}

\paragraph{Input representation and architecture.}
The reported feedforward model is the depth-9 union variant. For each object,
its input octree is the union of a conservatively rasterized depth-9 mesh
occupancy (with colors sampled from eight rendered albedo views) and a 200k
point cloud. The union is capped at 900,000 depth-9 cells; this cap limits the
encoder memory without changing the depth-7 coverage used by the renderer.
The octree uses \texttt{full\_depth}=4 and query depth 7. Its nine LPF input
channels encode position, within-cell offset, and local color. A sparse octree
U-Net maps the depth-9 input to 128 channels at query depth 7. These channels
are reshaped into eight 16-dimensional corner features per voxel and averaged
at shared corners, yielding the continuous vertex feature field used by the
renderer. The renderer has hidden width 128, four recurrent refinements,
8-dimensional voxel features, RMSNorm, and segment-position encoding. Its hit
threshold is 0.4, and the forward pass is chunked at 65,536 candidates (with a
maximum render batch of 262,144 queries).

Figure~\ref{fig:app_ff_unet} details the 3D U-Net used by the feedforward
encoder. The sparse octree input and point-cloud features enter a multiscale
encoder, are fused at the bottleneck, and are decoded with skip connections
to produce the output octree and point features. Each octree residual block
uses two $1\times1$ convolutions with RMSNorm and ReLU, followed by a residual
addition and a final ReLU, as shown in the lower panel.

\paragraph{Losses.}
The hit term uses balanced binary cross-entropy with positive weight 1.5,
Tversky weight 0.2 and $(\alpha,\beta,\gamma)=(0.3,0.7,1.0)$. Hole, floater,
and boundary penalties have weights 0.1, 0.1, and 0.2. Color uses weight 2.0,
a Charbonnier penalty, gradient weight 0.2, and masked SSIM weight 0.5. Normal
supervision uses cosine weight 0.1, gradient weight 0.3, and masked SSIM weight
0.3. We additionally use parameter $\ell_2$ regularization of $10^{-3}$,
ray-distance weight 0.5, final-stage weight 0.5, and detail scale 4.0.

\paragraph{Training and inference time.}
The reported d9-union model was selected on the full 1,000-object validation
set with a selection score of 1.3781. It was trained from random initialization
for three epochs (180,000 optimizer steps) on four RTX~3090 GPUs, corresponding
to approximately 50.6~h of uninterrupted training. Building the depth-9
occupancy cache is a separate one-time cost of approximately 5.2~h and 91~GB.
The 19.5~h figure in the main paper is the eight-GPU equivalent estimated from
measured per-iteration throughput. These training costs are not comparable to
the per-object timing table. Inference timing excludes data loading, rendering,
and metric evaluation.

\paragraph{Stored representation size.}
The shared feedforward network has 36.888M parameters and is not replicated per
object. Per-object state consists of depth-7 topology, a 16-dimensional fp16
vertex field, and an optional 8-dimensional fp16 voxel field. The 128-channel
U-Net output is an intermediate tensor and is not stored. With the voxel field
enabled, the seven reference objects occupy 381--1,674~KiB (806~KiB on
average); without it, they occupy 304--1,372~KiB (654~KiB on average).

\subsection{Complete Capacity Configurations for Table~1}
\label{app:table1_configurations}

\paragraph{NeuOctree (per-object optimization).}
For every cell, we optimize a geometry-only NeuOctree for 300 epochs over 242
training views, using two RTX~4090 GPUs with pair-balanced DDP. The optimizer
uses an initial learning rate of $2\times10^{-3}$ with linear decay to
$10^{-5}$. We use three decoder refinements, recurrent-pool sampling, no
voxel-feature branch, and fp16 storage. Each entry below is $D/F/H$ followed by
the measured payload. The feature dimension and decoder width are jointly
selected to fill the target budget; consequently, they vary across both
objects and rates. \Cref{tab:app_ours_configs} reports the complete settings.

\begin{table}[tbp]
\centering
\caption{NeuOctree configurations for the 28 cells in
\cref{tab:mesh_compression}; entries are $D/F/H$ and payload.}
\label{tab:app_ours_configs}
\scriptsize
\setlength{\tabcolsep}{2.5pt}
\renewcommand{\arraystretch}{1.13}
\begin{tabular}{@{}lcccc@{}}
\toprule
\textbf{Object} & \textbf{81 KiB} & \textbf{122 KiB} & \textbf{244 KiB} & \textbf{823 KiB} \\
\midrule
Armadillo & \shortstack{D6/F6/H54\\82,330 B} & \shortstack{D6/F9/H72\\124,822 B} & \shortstack{D7/F5/H81\\249,011 B} & \shortstack{D7/F16/H194\\840,881 B} \\
Buddha & \shortstack{D6/F9/H52\\82,812 B} & \shortstack{D6/F13/H71\\124,398 B} & \shortstack{D7/F6/H98\\249,359 B} & \shortstack{D7/F21/H194\\841,493 B} \\
Dragon & \shortstack{D6/F5/H55\\82,933 B} & \shortstack{D6/F8/H67\\124,699 B} & \shortstack{D6/F16/H102\\249,091 B} & \shortstack{D7/F13/H190\\841,674 B} \\
Einstein & \shortstack{D6/F5/H55\\82,589 B} & \shortstack{D6/F8/H68\\124,517 B} & \shortstack{D6/F16/H104\\248,765 B} & \shortstack{D7/F12/H180\\841,204 B} \\
Lucy & \shortstack{D6/F13/H55\\82,895 B} & \shortstack{D6/F19/H72\\124,357 B} & \shortstack{D7/F9/H101\\249,779 B} & \shortstack{D7/F31/H196\\842,165 B} \\
Metatron & \shortstack{D5/F11/H58\\82,599 B} & \shortstack{D5/F17/H71\\124,101 B} & \shortstack{D6/F8/H99\\249,026 B} & \shortstack{D7/F7/H175\\842,349 B} \\
XYZ-Dragon & \shortstack{D6/F11/H56\\82,801 B} & \shortstack{D6/F17/H69\\124,915 B} & \shortstack{D7/F9/H90\\249,004 B} & \shortstack{D7/F29/H195\\842,054 B} \\
\bottomrule
\end{tabular}
\end{table}

\paragraph{NGF.}
We keep the upstream NGF network fixed (20 features, eight Fourier-feature
levels, batch size 10, and a fixed seed); only the coarse quad mesh changes the
object-specific payload. Training uses four resolution levels, 100 iterations
per level, batch size 10, and 200 reference views on one RTX~4090. Each entry
in \cref{tab:app_ngf_configs} is requested faces $F$, realized coarse quads
$Q$, and measured payload $P$.
The paper metrics are recomputed with the common 16-view union-mask evaluator,
not NGF's native held-out-view evaluator.

\paragraph{QNDF.}
QNDF is run for 4,000 optimizer iterations on one RTX~4090. Its size is
controlled by the realized coarse-mesh face count $F$ and decoder width $H$.
Entries in \cref{tab:app_qndf_configs} are $F/H$ and measured payload. The
Dragon 823~KiB cell is an exception: a request for 1,900 faces produced a
remeshed surface with 12,798 faces and 6,381 vertices, explaining its larger
run time relative to the other cells at this rate.

\begin{table}[tbp]
\centering
\begin{minipage}[t]{0.49\linewidth}
\vspace{0pt}
\centering
\captionof{table}{NGF capacity configurations; entries are $F/Q$ and $P$.}
\label{tab:app_ngf_configs}
\resizebox{\linewidth}{!}{%
\begin{tabular}{@{}lcccc@{}}
\toprule
\textbf{Object} & \textbf{81 KiB} & \textbf{122 KiB} & \textbf{244 KiB} & \textbf{823 KiB} \\
\midrule
Armadillo & \shortstack{F197/Q281\\82,776 B} & \shortstack{F459/Q673\\124,928 B} & \shortstack{F1221/Q1830\\249,608 B} & \shortstack{F4881/Q7320\\842,528 B} \\
Buddha & \shortstack{F201/Q282\\82,700 B} & \shortstack{F461/Q677\\124,624 B} & \shortstack{F1237/Q1836\\249,704 B} & \shortstack{F4897/Q7331\\842,704 B} \\
Dragon & \shortstack{F227/Q270\\82,876 B} & \shortstack{F479/Q669\\124,864 B} & \shortstack{F1241/Q1832\\249,548 B} & \shortstack{F4897/Q7331\\842,704 B} \\
Einstein & \shortstack{F191/Q285\\82,748 B} & \shortstack{F451/Q675\\124,868 B} & \shortstack{F1221/Q1830\\249,608 B} & \shortstack{F4883/Q7320\\842,528 B} \\
Lucy & \shortstack{F203/Q279\\82,652 B} & \shortstack{F461/Q670\\124,696 B} & \shortstack{F1231/Q1826\\249,636 B} & \shortstack{F4905/Q7311\\842,476 B} \\
Metatron & \shortstack{F295/Q222\\82,476 B} & \shortstack{F505/Q651\\124,852 B} & \shortstack{F1235/Q1851\\249,852 B} & \shortstack{F4893/Q7338\\842,448 B} \\
XYZ-Dragon & \shortstack{F201/Q264\\82,688 B} & \shortstack{F454/Q620\\124,724 B} & \shortstack{F1219/Q1765\\249,672 B} & \shortstack{F4859/Q7227\\842,604 B} \\
\bottomrule
\end{tabular}}
\end{minipage}\hfill
\begin{minipage}[t]{0.49\linewidth}
\vspace{0pt}
\centering
\captionof{table}{QNDF capacity configurations; entries are coarse faces $F$,
decoder width $H$, and payload $P$.}
\label{tab:app_qndf_configs}
\resizebox{\linewidth}{!}{%
\begin{tabular}{@{}lcccc@{}}
\toprule
\textbf{Object} & \textbf{81 KiB} & \textbf{122 KiB} & \textbf{244 KiB} & \textbf{823 KiB} \\
\midrule
Armadillo & \shortstack{F1000/H24\\78,012 B} & \shortstack{F2700/H32\\121,532 B} & \shortstack{F4500/H64\\220,860 B} & \shortstack{F2700/H192\\824,764 B} \\
Buddha & \shortstack{F1000/H24\\77,820 B} & \shortstack{F2600/H32\\119,548 B} & \shortstack{F2600/H80\\233,724 B} & \shortstack{F2600/H192\\822,780 B} \\
Dragon & \shortstack{F1000/H24\\77,820 B} & \shortstack{F2600/H32\\119,548 B} & \shortstack{F2600/H80\\233,724 B} & \shortstack{F12798/H160\\791,868 B} \\
Einstein & \shortstack{F1000/H24\\77,884 B} & \shortstack{F2600/H32\\119,612 B} & \shortstack{F4210/H64\\233,788 B} & \shortstack{F2600/H192\\822,844 B} \\
Lucy & \shortstack{F983/H24\\77,756 B} & \shortstack{F2584/H32\\119,484 B} & \shortstack{F2584/H80\\233,660 B} & \shortstack{F2584/H192\\822,716 B} \\
Metatron & \shortstack{F1000/H24\\77,756 B} & \shortstack{F2600/H32\\119,484 B} & \shortstack{F4210/H64\\233,660 B} & \shortstack{F2600/H192\\822,716 B} \\
XYZ-Dragon & \shortstack{F1000/H24\\78,524 B} & \shortstack{F2600/H32\\120,380 B} & \shortstack{F2600/H80\\234,556 B} & \shortstack{F2600/H192\\823,612 B} \\
\bottomrule
\end{tabular}}
\end{minipage}
\end{table}

\paragraph{VQAD.}
We adapt VQAD to six-channel geometry supervision (normal and normalized
position), use four LODs, and optimize for 100 epochs over 242 training views
with RMSprop at $10^{-3}$ and AMP. Table~1 uses the actual evaluated
configuration for each cell: two 81~KiB cells use the final 3-bit setting, 13
further cells use uniform splitting, and the remaining cells use the settings
listed in \cref{tab:app_vqad_configs}. Entries are $L/B/F/H/K$, where $K$ is
the number of decoder layers, followed by payload $P$.

\paragraph{N-BVH.}
N-BVH uses four features per level, eight levels, 64 neurons, four hidden
layers, \texttt{WorldT} depth learning, a last split at iteration 3,000, split
and split-interval scale factors of 2, and 10,000 training steps on one
RTX~4090. The hash-map sizes are $2^9$, $2^9$, $2^{10}$, and $2^{11}$ across
the four rates. Capacity is controlled by split intervals 235, 100, 60, and
28, producing the node counts in \cref{tab:app_nbvh_configs}. Entries are node
count $N$ and payload $P$.

\begin{table}[tbp]
\centering
\begin{minipage}[t]{0.49\linewidth}
\vspace{0pt}
\centering
\captionof{table}{VQAD configurations used for Table~1; entries are
$L/B/F/H/K$ and payload.}
\label{tab:app_vqad_configs}
\resizebox{\linewidth}{!}{%
\begin{tabular}{@{}lcccc@{}}
\toprule
\textbf{Object} & \textbf{81 KiB} & \textbf{122 KiB} & \textbf{244 KiB} & \textbf{823 KiB} \\
\midrule
Armadillo & \shortstack{L7/B3/F16/H64/2\\77,330 B} & \shortstack{L8/B2/F16/H64/2\\115,696 B} & \shortstack{L9/B2/F16/H128/1\\246,108 B} & \shortstack{L9/B4/F16/H256/2\\754,571 B} \\
Buddha & \shortstack{L8/B2/F16/H64/1\\72,862 B} & \shortstack{L8/B4/F16/H64/1\\110,178 B} & \shortstack{L9/B4/F16/H64/2\\243,210 B} & \shortstack{L9/B4/F16/H256/2\\639,034 B} \\
Dragon & \shortstack{L7/B3/F16/H64/2\\75,326 B} & \shortstack{L8/B2/F16/H64/2\\113,041 B} & \shortstack{L9/B2/F16/H64/2\\231,539 B} & \shortstack{L9/B4/F16/H256/2\\756,808 B} \\
Einstein & \shortstack{L8/B2/F16/H64/1\\74,639 B} & \shortstack{L8/B4/F16/H64/1\\113,129 B} & \shortstack{L9/B4/F16/H64/1\\237,149 B} & \shortstack{L9/B4/F16/H256/2\\650,029 B} \\
Lucy & \shortstack{L8/B2/F16/H64/1\\75,328 B} & \shortstack{L8/B4/F16/H64/1\\114,267 B} & \shortstack{L9/B4/F16/H64/1\\239,716 B} & \shortstack{L9/B4/F16/H256/2\\652,339 B} \\
Metatron & \shortstack{L7/B3/F16/H64/1\\73,554 B} & \shortstack{L7/B4/F16/H128/1\\122,588 B} & \shortstack{L8/B4/F16/H64/2\\232,907 B} & \shortstack{L9/B4/F16/H256/1\\690,683 B} \\
XYZ-Dragon & \shortstack{L8/B2/F16/H64/1\\81,864 B} & \shortstack{L8/B3/F16/H64/2\\119,362 B} & \shortstack{L9/B3/F16/H64/2\\246,025 B} & \shortstack{L9/B4/F16/H256/2\\698,100 B} \\
\bottomrule
\end{tabular}}
\end{minipage}\hfill
\begin{minipage}[t]{0.49\linewidth}
\vspace{0pt}
\centering
\captionof{table}{N-BVH capacity configurations. The calibrated node count is
shared across objects at a given rate.}
\label{tab:app_nbvh_configs}
\resizebox{\linewidth}{!}{%
\begin{tabular}{@{}lcccc@{}}
\toprule
\textbf{Object} & \textbf{81 KiB} & \textbf{122 KiB} & \textbf{244 KiB} & \textbf{823 KiB} \\
\midrule
Armadillo & \shortstack{N219\\81,535 B} & \shortstack{N1,243\\122,501 B} & \shortstack{N3,675\\248,457 B} & \shortstack{N16,475\\817,805 B} \\
Buddha & \shortstack{N219\\81,693 B} & \shortstack{N1,243\\122,660 B} & \shortstack{N3,675\\248,616 B} & \shortstack{N16,475\\817,964 B} \\
Dragon & \shortstack{N219\\81,693 B} & \shortstack{N1,243\\122,660 B} & \shortstack{N3,675\\248,616 B} & \shortstack{N16,475\\817,964 B} \\
Einstein & \shortstack{N219\\81,695 B} & \shortstack{N1,243\\122,662 B} & \shortstack{N3,675\\248,618 B} & \shortstack{N16,475\\817,966 B} \\
Lucy & \shortstack{N219\\81,691 B} & \shortstack{N1,243\\122,658 B} & \shortstack{N3,675\\248,614 B} & \shortstack{N16,475\\817,962 B} \\
Metatron & \shortstack{N219\\81,695 B} & \shortstack{N1,243\\122,662 B} & \shortstack{N3,675\\248,618 B} & \shortstack{N16,475\\817,966 B} \\
XYZ-Dragon & \shortstack{N219\\81,697 B} & \shortstack{N1,243\\122,664 B} & \shortstack{N3,675\\248,620 B} & \shortstack{N16,475\\817,968 B} \\
\bottomrule
\end{tabular}}
\end{minipage}
\end{table}

\subsection{Training Hardware and Runtime}
\label{app:runtime}

All per-object timing runs were executed on the same RTX~4090 server. Ours
uses two GPUs with DDP; each baseline uses one GPU. The values in
\cref{tab:app_runtime} are wall-clock minutes per object, averaged over all
seven meshes. They are practical end-to-end fitting costs under each method's
own training budget, rather than equal-GPU or equal-step throughput
measurements. The N-BVH times exclude the separate capacity-calibration probe.
The VQAD timing sweep uses a uniform 65,536-sample protocol for all 28 cells;
its timing configurations may therefore differ from the quality configurations
in \cref{tab:app_vqad_configs}.

\begin{table}[H]
\centering
\caption{Measured fitting time and hardware. Times are minutes per object.}
\label{tab:app_runtime}
\scriptsize
\setlength{\tabcolsep}{2.7pt}
\renewcommand{\arraystretch}{1.08}
\begin{tabular}{@{}llccccc@{}}
\toprule
\textbf{Method} & \textbf{Training budget} & \textbf{GPU} & \textbf{81} & \textbf{122} & \textbf{244} & \textbf{823} \\
\midrule
NGF & $4\times100$ iterations & $1\times4090$ & 3.9 & 4.1 & 6.3 & 19.1 \\
QNDF & 4,000 iterations & $1\times4090$ & 14.5 & 41.0 & 42.4 & 53.2 \\
VQAD & $100\times242$ steps & $1\times4090$ & 16.0 & 35.2 & 36.7 & 51.8 \\
N-BVH & 10,000 steps & $1\times4090$ & 3.5 & 3.7 & 3.6 & 3.1 \\
Ours & $300\times121$ steps & $2\times4090$ & 28.2 & 30.2 & 31.1 & 37.7 \\
\bottomrule
\end{tabular}
\end{table}

Feedforward training uses four RTX~3090 GPUs. The capacity and timing
experiments use the GPU allocations reported in \cref{tab:app_runtime}, and
the generation experiments use four RTX~5090 GPUs. We report only the devices
allocated to each experiment rather than the total capacity of the host.

\subsection{Native 3D Generation Details}
\label{app:generation_details}

The generation demonstration uses NeuOctree itself as the output domain. A
sparse-octree VAE encodes the occupancy and feature tensors of a renderable
NeuOctree into an $8^3$ latent grid. An image-conditioned flow-matching DiT
predicts both latent features and occupancy; the VAE decoder then reconstructs
the octree feature field, which can be queried by the same surface renderer as
the reconstructed assets. No mesh extraction or post-hoc texture baking is
required, and the paper does not use this experiment for a quantitative
comparison with external generation systems.

The generation setup uses 5,000 ShapeNet objects from ten categories.
The VAE receives a $128^3$ representation, downsamples it four times to $8^3$,
and uses a latent dimension of 32. It is trained with AdamW (learning rate
$10^{-4}$, weight decay 0.01), batch size 32, on four RTX~5090 GPUs. For the
image-conditioned model, 16 random Blender views are rendered per asset, with
one retained for testing. The DiT has approximately 0.7B parameters: width
1,536, 16 transformer blocks, 12 attention heads, and MLP width 8,192. It uses
the same optimizer settings, batch size 16, and four RTX~5090 GPUs.

\subsection{Reproducibility Notes and Limitations}
\label{app:reproducibility_notes}

The capacity tables report measured payloads rather than inferring size from a
compression-ratio label. The actual capacity controls are the values in
\cref{tab:app_ours_configs,tab:app_ngf_configs,tab:app_qndf_configs,tab:app_vqad_configs,tab:app_nbvh_configs}.
The VQAD quality table uses the tested configuration listed for each cell,
whereas its timing table uses a uniform 28-cell protocol.
QNDF's Dragon 823~KiB remeshing exception and N-BVH's excluded calibration
probe are disclosed above. Finally, the feedforward inference time measures
representation construction only; image rendering and evaluation are separate
downstream operations.

\subsection{Additional Qualitative Results}
\label{app:additional_qualitative}

\begin{figure}[H]
\centering
\includegraphics[width=0.82\linewidth]{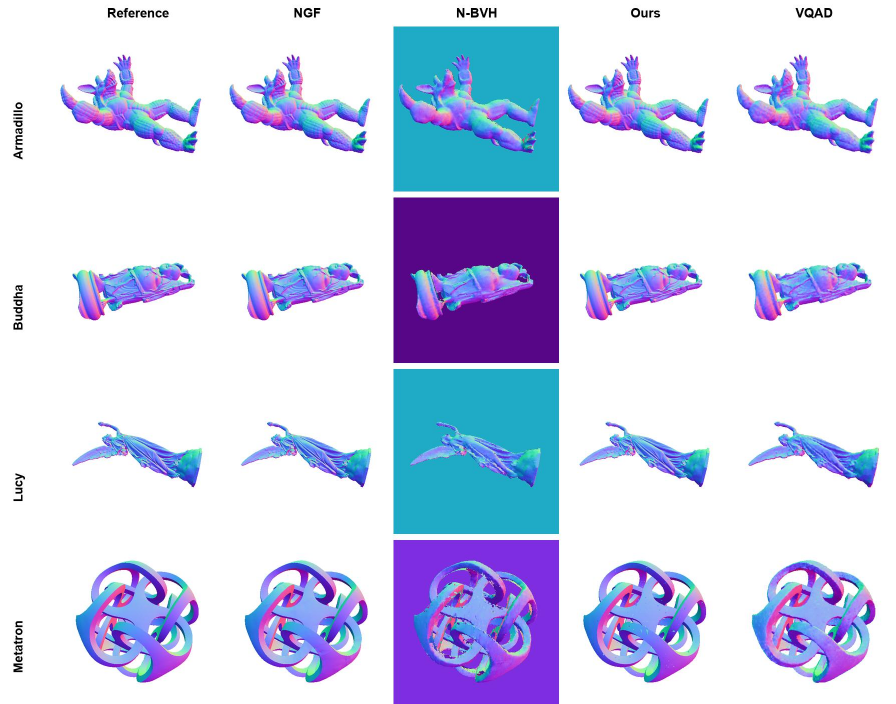}
\caption{Additional 823~KiB normal reconstructions for Armadillo, Buddha,
Lucy, and Metatron. Columns show the reference, NGF, N-BVH, NeuOctree, and
VQAD under matched views.}
\label{fig:app_remaining_models}
\vspace{4pt}
\includegraphics[width=0.44\linewidth]{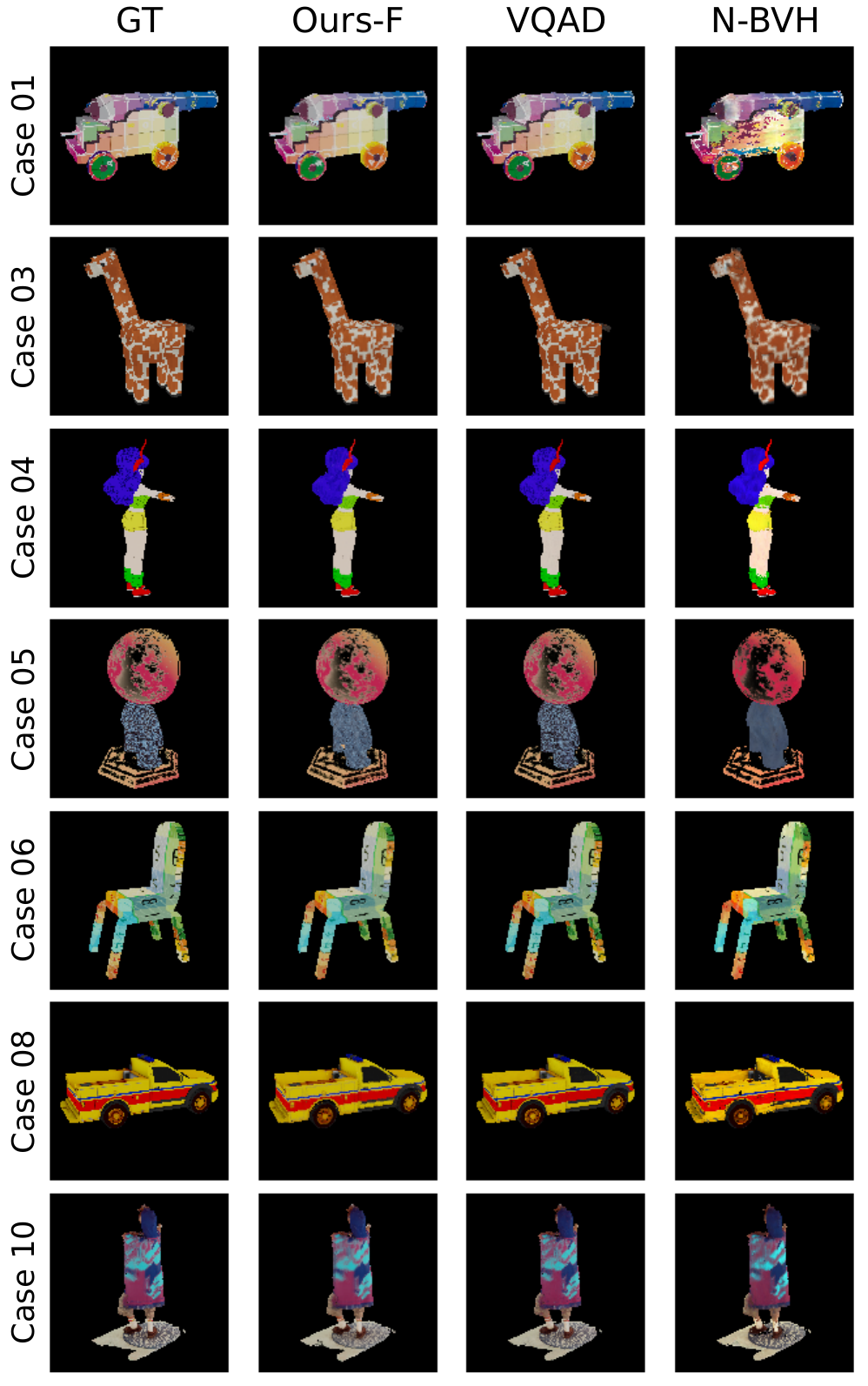}
\caption{Additional Objaverse albedo comparisons at a nominal 823~KiB
per-object budget. NeuOctree-F uses one shared feedforward model.}
\label{fig:app_objaverse_albedo}
\end{figure}

\Cref{fig:app_remaining_models} supplies the four Table~1 objects not shown in
the main qualitative compression figure. The N-BVH render uses its method
background, which is retained rather than cosmetically altered.

\Cref{fig:app_objaverse_albedo} shows seven further Objaverse objects under a
matched view. NeuOctree-F preserves large color regions and thin structures
without per-object fitting; VQAD is often similarly sharp, whereas N-BVH more
frequently exhibits color fragmentation or smoothing at the same nominal
823~KiB budget.

